\documentclass[pre,twocolumn,aps,showpacs,floatfix]{revtex4}

\usepackage{graphicx}
\usepackage[latin1]{inputenc}
\usepackage{dcolumn}
\usepackage{amsmath}
\usepackage{amsfonts}
\usepackage{amssymb}
\usepackage{epsf}
\usepackage{epsfig}
\usepackage{bm}

\begin{document}
\title{Coagulation reactions in low dimensions: Revisiting subdiffusive $A+A$ reactions in one dimension}
\author{S. B. Yuste$^{1}$, J. J. Ruiz-Lorenzo$^{1}$, and Katja Lindenberg$^{2}$}
\affiliation{$^{(1)}$
Departamento de F\'{\i}sica, Universidad de Extremadura,
E-06071 Badajoz, Spain\\
$^{(2)}$Department of Chemistry and Biochemistry,
University of California San Diego, 9500 Gilman Drive, La Jolla, CA
92093-0340, USA}

\begin{abstract}
We present a theory for the coagulation reaction $A+A\to A$ for particles
moving subdiffusively in one dimension. Our theory is tested against
numerical simulations of the concentration of $A$ particles as a function of
time (``anomalous kinetics'') and of the interparticle distribution function as
a function of interparticle distance and time.  We find that the theory captures
the correct behavior asymptotically and also at early times, and that it does so
whether the particles are nearly diffusive or very subdiffusive. We find that,
as in the normal diffusion problem, an interparticle gap responsible for the
anomalous kinetics develops and grows with time.  This corrects an earlier
claim to the contrary on our part.
\end{abstract}

\pacs{02.50.Ey,82.40.-g,82.33.-z,05.90.+m}

\maketitle

\section{Introduction}
\label{sec:intro}
The traditional laws of mass action that describe the time evolution of
the macroscopic global concentrations of reactants and products in chemical
reactions assume that the system is well stirred and therefore spatially
homogeneous.  However, there are many situations when a
reactive system is not well mixed; in that case one must
deal with local concentrations and account for the effects of spatial
inhomogeneities on local reaction rates.  Spatial variations
in concentrations of reactants lead to changes (often called ``anomalies")
in the time dependences of the spatially averaged macroscopic
concentrations.  One often encounters this situation when diffusion is
the only mixing mechanism, particularly when diffusion is the rate
limiting step for a reaction to occur.  The inefficiency of diffusion as
a mixing mechanism becomes more pronounced with decreasing
dimensionality of the system, and it is therefore commonly accepted that
diffusion-limited reactions in constrained geometries exhibit kinetic
``anomalies"~\cite{Kotomin}.
An exact description of diffusion-limited reactions in the face of
nonuniform spatial distributions of reactants typically requires an
infinite hierarchy of correlation functions to properly incorporate
spatial correlations.  In practice, such hierarchies are often truncated
at the first or second level, giving rise to well-known
reaction-diffusion equations whose solutions are sometimes fairly
satisfactory (and sometimes not) in capturing the principal deviations
form the laws of mass action~\cite{Something1,Something2}.

The judgment as to the success or failure of approximate
reaction-diffusion models has mostly relied on comparisons with
numerical simulations.  There has always been a quest for exact
analytic solutions against which approximate reaction-diffusion theories
could be tested, but there have been very few successes, among them
the coagulation ($A+A\to A$ and $A+A\rightleftharpoons A$) and
annihilation ($A+A\to 0$) reactions~\cite{Something2,Something3}.  
Exact solutions in these cases have
been possible because, if instead of focusing on the concentration of
reactants, one focuses on the evolution of empty intervals (coagulation
reactions) or on the number parity (even or odd) of particles in an
interval (annihilation reactions), one arrives at \emph{exactly linear}
diffusion equations.  These solutions have provided a wealth
of information against which to measure the approximate solutions
obtained from more standard approaches for these particular reactions.
Unfortunately, the interval approaches are not generalizable to other
reactions.

In the past few years there have also been attempts to understand
chemical reactions in low dimensions when the reactants move
\emph{subdiffusively}.  This has been a particularly interesting subject
in view of the many systems, mainly biological, in which such reactions
occur in a complex or constrained environment that does not even permit
ordinary diffusive motion of chemical species.
All the difficulties encountered in diffusive systems are exacerbated
in this case because motions are even slower, a fact that turns out to
have profound consequences on spatial as well as temporal
correlations~\cite{ourlatest}
and to cause much greater difficulties in both numerical simulations and
analytic attempts.  One approach to this problem has
been to essentially adapt the existing reaction-diffusion models by
modifying the diffusive description to a subdiffusive one involving
fractional diffusion operators.  These descriptions are
phenomenological in nature, and much work remains to be done to
understand how one might arrive at a mesoscopic description from
microscopic considerations.  In particular, there are many questions
about how to describe the reaction terms in subdiffusive systems, and
there is clear evidence that the problem is usually not even separable
into a simple sum of a term describing the motion of the
particles and one describing the reaction.

The subdiffusive coagulation and annihilation problems seemed to offer a
parallel opportunity for exact solution if one again concentrated on the
properties of intervals.  We followed this route in our earlier
work~\cite{ourold}, and
formulated what we thought to be exact subdiffusive equations for the
intervals from whose solution one could calculate the particle
concentrations and interparticle separations, as had been done in the
diffusive problem.  These solutions led us to the conclusion that
interparticle gaps do not occur in the subiffusive problem, a result
that seemed reasonable for particles that move sufficiently slowly.
However, subsequent numerical simulations indicated that a gap
\emph{does}
develop no matter how subdiffusive the particles, and this led to a
reassessment of the assumptions of our original theory.  In this paper
we present this numerical evidence, a description of the difficulty with
the original theory, and a new theory which, albeit still approximate,
seems to capture the correct behavior to a very high degree of accuracy.
Also, we recently developed a mean field theory for this problem~\cite{ourlatest}
but it is only valid for dimension 3 and higher. 
We concentrate on the coagulation problem, although a similar approach
may be helpful for the annihilation reaction.

In Sec.~\ref{traditionaltheory} we describe the difficulties with our previous
1D
theory~\cite{ourold}. Section~\ref{newtheory} presents our new theory.
Section~\ref{conundrum} discusses some
special interesting issues that arise in this problem, and in
Sec.~\ref{numerical} we present the numerical
evidence to support our theory. We conclude with a summary in
Sec.~\ref{conclusions}.

\section{Traditional Theory}
\label{traditionaltheory}
The interparticle distribution function method focuses on the ``empty
interval" function $E(x,t)$, defined as the probability that an interval
of length $x$ is empty of particles at time $t$.  From $E(x,t)$ one
obtains the concentration of particles,
\begin{equation}
c(t) = - \left. \frac{\partial E(x,t)}{\partial x}\right|_{x=0},
\label{c}
\end{equation}
and the interparticle distribution function,
\begin{equation}
p(x,t) = \frac{1}{c(t)} \frac{\partial^2 E(x,t)}{\partial x^2},
\label{p}
\end{equation}
which is the probability density that the first $A$ to be found to one side of
a given $A$ at time $t$ is a distance $x$ away.
The question of interest is how to determine the empty interval function.

If the particles undergo normal diffusion, the probability
density for finding a particle $A$ at $y$ at time $t$ in the absence
of reactions obeys the diffusion equation
\begin{equation}
\frac{\partial}{\partial t} P(y,t) = D\frac{\partial^2}{\partial y^2}
P(y,t).
\label{diff}
\end{equation}
In the presence of the coagulation reaction, the empty interval
function focuses on the diffusive motion of the
particles at each end of an empty interval and one readily
arrives at the exact equation
\begin{equation}
\frac{\partial}{\partial t} E(x,t) = 2D\frac{\partial^2}{\partial x^2}
E(x,t).
\label{diffint}
\end{equation}
The equation is easily understood from the fact that the empty interval dynamics
is the same as that of the individual particles but with double the diffusion
coefficient to reflect the relative motion of two diffusive particles.
This readily tractable equation, together with appropriate boundary
conditions, exactly solves the diffusive $A+A\to A$ problem in one
dimension.

A standard approach for the description of subiffusive processes starts from
the continuous time random
walk (CTRW) formalism, in which a walker jumps from one site on a lattice to
another in consecutive steps as
time proceeds~\cite{MetKlaPhysRep,Hughes1}. 
Both the jump distances $n$ and times $t$ between jumps are random variables
drawn from a
probability distribution function $\phi(n,t)$.  If the jump distances and jump
times are independent
random variables, then this distribution function is simply a product,
$\phi(n,t) = w(n)\psi(t)$. 
''Normal`` CTRWs are associated with distributions $w(n)$ and $\psi(t)$
that have finite first
moments, and the scaling limit that leads from the random walk to the diffusion
equation is well
known.  One way to obtain a subdiffusive process is for the
waiting time distribution $\psi(t)$ to be heavy-tailed, i.e., $\psi(t)\sim
t^{-1-\gamma}$
with $0<\gamma<1$ for long times, so that the mean waiting time between jumps
diverges.
In this case a number of scaling approaches can be found in the literature, with a particularly
helpful discussion in~\cite{Scalas}.  In the absence of reactions,
in the continuum limit with a particular scaling one arrives at a ``fractional diffusion
equation" for the evolution of the probability density $P(y,t)$ of a subdiffusive particle, 
\begin{equation}
\frac{\partial}{\partial t} P(y,t) =
~_{0}D_{t}^{1-\alpha}K_\alpha \frac{\partial^2}{\partial y^2} P(y,t),
\label{subdiff}
\end{equation}
where $~_{0}D_{t}^{1-\alpha}$ is the Riemann-Liouville operator,
\begin{equation}
~_{0}D_t^{1-\alpha} P(y,t) =\frac{1}{\Gamma(\alpha)}
\frac{\partial}{\partial t} \int_0^t d\tau
\frac{P(y,\tau)}{(t-\tau)^{1-\alpha}},
\end{equation}
and $K_\alpha$ is the generalized diffusion coefficient.  The mean
square displacement of the $A$ particle for large $t$ that follows from
this evolution equation is
\begin{equation}
\langle y^2(t)\rangle \sim \frac{2K_\alpha}{\Gamma(1+\alpha)} t^\alpha,
\label{msd}
\end{equation}
which reduces to the ordinary diffusion result when $\alpha =1$ ($K_1=D$).
In our earlier work we argued that the same reasoning that led
from Eq.~(\ref{diff}) to Eq.~(\ref{diffint}) would also lead from
Eq.~(\ref{subdiff}) to the interval equation
\begin{equation}
\frac{\partial}{\partial t} E(x,t) =
~_{0}D_{t}^{1-\alpha}2K_\alpha \frac{\partial^2}{\partial x^2} E(x,t),
\label{intsubdiff}
\end{equation}
and then calculated the particle concentration and interparticle
distribution function from the solution of this equation with
appropriate boundary conditions.

The difficulty with this reasoning is the fact that in the subdiffusive
problem we must face the issue of \emph{aging}~\cite{Barkai}, which we have not
done above.  To describe the issue, we again turn to a CTRW point
of view of the problem. Consider first the diffusion problem, where
the continuum limit leads to a diffusion equation.  In this limit,
in which the mean time
between steps and the step length go to zero in an appropriate way, one
arrives at the diffusion equation.  In particular, in the diffusive
problem the steps are sufficiently frequent that one need not keep
track of extremely long sojourns of a particle at any one site, and what
transpired in the past is quickly forgotten.  A similar formulation of the
\emph{subdiffusive} problem involves waiting time distributions with an
infinite mean time between steps.  Even worse, because of aging, the waiting time distribution for a
particle to take its next step at time $t$ having arrived at its current location at time $t'$
is no longer simply a function of the difference $t-t'$ but of each time separately.
To understand the effects of these complications,
suppose that an observer looks at the system at time $t$ and sees an empty
interval of size $x$ at that instant. For normally diffusive
particles, the evolution of the size of this interval does not depend on the
time at which the interval was first created. However, in order to predict
the evolution of this gap for subdiffusive particles the observer must
know how long each of the two particles at the ends of this interval have been
at that location. Due to aging, the evolution depends on the times $t-t_l$ and
$t-t_r$, where $t_l$ and $t_r$ are the times at which the left and right
particles jumped to the locations seen by the observer at time $t$.
Moreover, even if the two particles arrived at this location at the
same time, i.e., $t_l=t_r$, the shortening and lengthening of this gap is not
correctly described by Eq.~(\ref{intsubdiff}),as
we will show in Sec.~\ref{newtheory}
[cf. Eqs.~\eqref{propagator}-\eqref{3.17}].

\section{Theory Revisited}
\label{newtheory}
To find a more appropriate description for the evolution of the empty
interval, we introduce two hypotheses.  As we state
each hypothesis, we provide an argument as to its approximate nature.

We start by forgetting the reaction for a moment, and simply consider
the motion of (mutually transparent) $A$ particles on an infinite line.
We define ${\mathcal P}_\alpha(x,t)$ as the probability density for $x$
to be the distance between two subdiffusive particles at time $t$
with the initial condition ${\mathcal P}_\alpha(x,0) = \delta(x)$, i.e.,
\begin{equation}
\mathcal{P}_\alpha(x,t)=\int_{-\infty}^\infty dy P_\alpha(x-y,t)
P_\alpha(y,t).
\label{propagator}
\end{equation}
Here $P_\alpha(y,t)$ is the propagator of the subdiffusion equation,
i.e., the solution of Eq.~(\ref{subdiff}) with the initial condition
$P_\alpha(y,0)=\delta(y)$. In Fourier space the solution in closed form
is the Mittag-Leffler function,
\begin{equation}
\tilde P_\alpha(q,t)=E_\alpha(-K_\alpha q^2 t^\alpha).
\label{ml}
\end{equation}
The Fourier transform of Eq.~(\ref{propagator}) then tells us that
\begin{equation}
       \mathcal{\tilde  P}_\alpha(q,t)=\left(\tilde
P_\alpha(q,t)\right)^2=\left[E_\alpha(-K_\alpha q^2 t^\alpha)\right]^2,
\label{solution}
\end{equation}
and correspondingly,
\begin{eqnarray}
\mathcal{P}_\alpha(y,t) &=&
\frac{1}{2\pi} \int_{-\infty}^\infty dq
e^{-iqy}\left[E_\alpha(-K_\alpha q^2 t^\alpha)\right]^2 \nonumber\\
&=& \frac{1}{\pi} \int_0^\infty dq
\cos(qy)\left[E_\alpha(-K_\alpha q^2 t^\alpha)\right]^2.
\label{green}
\end{eqnarray}

We do not know the evolution equation for
${\mathcal P}_\alpha(x,t)$, i.e., we do not know the operator
$\mathfrak{F}_\alpha$ such that $ \mathfrak{F}_\alpha {\mathcal
P}_\alpha(x,t)=0$, {\em except for} $\alpha=1$.  In this case the
Mittag-Leffler function reduces to an exponential, $E_1(-z) = \exp(-z)$,
so that $\mathcal{\tilde  P}_1(q,t)=\exp(-2 K_1 q^2 t)$ and
the corresponding function $\mathcal{P}_1(x,t)$ is the solution
of the diffusion equation~(\ref{diff}) but with double the diffusion
coefficient,
\begin{equation}
\frac{\partial}{\partial t} \mathcal{P}_1(x,t) =
2D\frac{\partial^2}{\partial x^2}
\mathcal{P}_1(x,t),
\end{equation}
with $K_1\equiv D$.  However, for $\alpha<1$ the Fourier transform of
the solution of the subdiffusion equation with double the subdiffusion
coefficient,
\begin{equation}
\frac{\partial }{\partial t} \mathcal{Q}_\alpha(x,t)\; = \;
_0D_{t}^{1-\alpha }\;
2K_\alpha \frac{\partial^2}{\partial x^2} \mathcal{Q}_\alpha(x,t),
\label{ecuDifudis}
\end{equation}
is the Mittag-Leffler function of twice the argument in Eq.~(\ref{ml}),
\begin{equation}
\tilde {\mathcal{Q}}_\alpha(q,t)=E_\alpha(-2K_\alpha q^2 t^\alpha).
\end{equation}
Clearly, except for $\alpha=1$,
\begin{equation}
\left[E_\alpha(-K_\alpha q^2 t^\alpha)\right]^2\neq E_\alpha(-2 K_\alpha
q^2 t^\alpha),
\end{equation}
so that $\mathcal{P}_\alpha\neq \mathcal{Q}_\alpha$, and the operator
$\mathfrak{F}_\alpha$ is not straightforwardly obtained from the
subdiffusion equation,
\begin{equation}
\label{3.17}
\mathfrak{F}_\alpha \neq  \frac{\partial }{\partial t}
\;-\; _0D_{t}^{1-\alpha }\;
2K_\alpha \frac{\partial^2}{\partial x^2}.
\end{equation}

\subsection{Hypothesis 1}
\label{hypothesis1}
The central hypothesis of our new theory is that \emph{the (unknown)
equation that describes the evolution of}
${\mathcal P}_\alpha(x,t)$ \emph{is the same as the equation that
describes the evolution of the empty interval function} $E(x,t)$, i.e.,
that
\begin{equation}
\mathfrak{F}_\alpha E(x,t) =0, \qquad 0\leq x <\infty
\end{equation}
This hypothesis is in general an approximation because the
distribution of particles is not expected to
be the same in the presence and absence of the reaction (although in higher dimensions it
is~\cite{ourlatest}). While there is
no memory of prior reaction events in the diffusive problem, in the
subdiffusive case this memory persists and may lead to a
distortion of the distribution relative to that of the particles when there is no reaction.
We only know how to assess the severity of this approximation via
comparison of results with numerical simulations~\cite{ourlatest} (see
Sec.~\ref{numerical}).

How does this hypothesis help us find the empty interval distribution
when in fact we do not know the operator $\mathfrak{F}_\alpha$? It helps
us because we know the solution ${\mathcal P}_\alpha(x,t)$, which is
then the Green function or propagator for the empty interval
distribution.  In other words, while we do not know the equation of
evolution for $E(x,t)$, we have sufficient information to construct
the function $E(x,t)$ itself explicitly.  The interval function is subject
to the additional conditions
\begin{equation}
\label{EcuE}
\begin{aligned}
E(0,t)&=1,\\
E(\infty,t)&=0\\
E(x,0)&=f(x).
\end{aligned}
\end{equation}
The first boundary condition is the probability that an interval of
vanishing width is empty.  As in the diffusive problem, this probability
must be unity. The second simply recognizes the existence of particles
at any time $t$, and the third is determined by the initial distribution
of empty intervals.

In order to construct the solution $E(x,t)$ from our knowledge of the
propagator $\mathcal{P}_\alpha(x,t)$, we introduce our second
hypothesis.

\subsection{Hypothesis 2}
\label{hypothesis2}
The second hypothesis of our theory is that \emph{the operator}
$\mathfrak{F}_\alpha$ \emph{is linear}. We know this operator to
be linear when $\alpha=1$, but it is most likely an
approximation when $\alpha<1$, although our lack of information about
$\mathfrak{F}_\alpha$ makes it difficult to assess.  While evolution
equations describing various quantities in some other reaction-subdiffusion
problems are in fact not linear, the connection between those systems
and the one considered here is not clear.

In any case, under this hypothesis we can split $E(x,t)$ into two parts,
each of which satisfies conditions whose symmetry properties allow for
the advantageous use of our knowledge of the propagator.
In particular, we write
\begin{equation}
E(x,t) = E^{\mathcal T}(x,t) +E^{\mathcal A}(x,t),
\end{equation}
where the individual pieces satisfy the following:
\begin{equation}
\label{EcuEa}
\begin{aligned}
\mathfrak{F}_\alpha E^{\mathcal T}(x,t)&=0, \qquad 0\leq x <\infty\\
E^{\mathcal T}(0,t)&=0,\\
E^{\mathcal T}(\infty,t)&=0\\
E^{\mathcal T}(x,0)&=f(x)\\
\end{aligned}
\end{equation}
and
\begin{equation}
\label{EcuEb}
\begin{aligned}
\mathfrak{F}_\alpha E^{\mathcal A}(x,t)&=0, \qquad 0\leq x <\infty\\
E^{\mathcal A}(0,t)&=1,\\
E^{\mathcal A}(\infty,t)&=0\\
E^{\mathcal A}(x,0)&=0.
\end{aligned}
\end{equation}
The superscripts ${\mathcal T}$ and ${\mathcal A}$ denote ``transient"
and ``asymptotic," respectively, for reasons that become evident below.

\subsection{Asymptotic (Long Time) Results}

In general, the solution $E(x,t)$ and its derivative observables depend
on the initial distribution $f(x)$.  However, since the evolution of the
interval function is in some sense necessarily subdiffusion-like, we
expect that the given boundary conditions lead to a decay of
$E^{\mathcal T}(x,t)$,
i.e., $E^{\mathcal T}(x,t)\to 0$ as $t\to\infty$, while
$E^{\mathcal A}(x,t)$ can not decay.
Thus at sufficiently long times the behavior of the interval function
will be dominated by that of $E^{\mathcal A}$.  One must keep in mind
that the decay of $E^{\mathcal T}$ may be slow (especially when comparing
with numerical simulations).  In particular, whereas diffusive modes decay
exponentially, subdiffusive modes typically decay only as $t^{-\alpha}$
for $\alpha <1$.  Nevertheless, this provides a helpful element if one
is interested in the asymptotic behavior because
$E^{\mathcal A}(x,t)$ does not depend on the initial distribution and can
therefore be pursued once and for all.  We therefore focus on it first.

The solution for $E^{\mathcal A}(x,t)$ can be obtained by the
method of images.
For this purpose, we consider the related problem
\begin{equation}
\label{EcuIma}
\begin{aligned}
\mathfrak{F}_\alpha \mathcal{E}^{\mathcal A}(x,t)&=0, \qquad -\infty<x <\infty  \\
\mathcal{E}^{\mathcal A}(-\infty,t)&=2,\\
\mathcal{E}^{\mathcal A}(\infty,t)&=0\\
\mathcal{E}^{\mathcal A}(x,0)&=2-2\theta(x),
\end{aligned}
\end{equation}
where $\theta(x)$ is the Heaviside step function.  The symmetry of the
problem immediately leads to the conclusion that
$\mathcal{E}^{\mathcal A}(0,t)=1$.
Therefore the solution of this problem for $x\ge 0$ is just the solution
$E^{\mathcal A}(x,t)$ of Eq.~\eqref{EcuEb}, that is,
$\mathcal{E}^{\mathcal A}(x,t)=E^{\mathcal A}(x,t)$ for $x\ge 0$.
On the other hand, we can write
\begin{equation}\label{}
   \mathcal{E}^{\mathcal A}(x,0)=2-2\theta(x)=2\int_{-\infty}^0 dy \delta(x-y).
\end{equation}
Since we know that the Green function of $\mathfrak{F}_\alpha$ is
$\mathcal{P}_\alpha(x,t)$, and since we assume that
$\mathfrak{F}_\alpha$ is linear, the solution of Eq.~\eqref{EcuIma} is
\begin{equation}\label{}
\mathcal{E}^{\mathcal A}(x,t)=2\int_{-\infty}^0 dy \mathcal{P}_\alpha(x-y,t)
\end{equation}
or, equivalently,
\begin{equation}\label{}
\mathcal{E}^{\mathcal A}(x,t)=2\int_{x}^\infty dy \mathcal{P}_\alpha(y,t) .
\end{equation}
Therefore
\begin{equation}\label{Ebxt}
E^{\mathcal A}(x,t)=2\int_{x}^\infty dy \mathcal{P}_\alpha(y,t)  \qquad
\text{for}\quad x \ge 0,
\end{equation}
with $\mathcal{P}_\alpha(y,t)$ given in Eq.~(\ref{green}).
It is noteworthy that $E^{\mathcal A}(x,t)$ depends on $x$ and $t$ only via the
variable 
\begin{equation}
z=\frac{c_\alpha}{\sqrt{K_\alpha t^\alpha}} x, 
\end{equation}
where
\begin{equation}
c_\alpha\equiv \frac{2}{\pi } \int_{0}^\infty dy
\left[E_\alpha(-y^2)\right]^2.
\label{calpha}
\end{equation}
The changes of variables $w\equiv \sqrt{K_\alpha t^\alpha}q$ and
$u\equiv (c_\alpha/ \sqrt{K_\alpha t^\alpha}) y$ immediately lead to
\begin{eqnarray}
\label{Ez}
E^{\mathcal A}(x,t)&=&E(z)\nonumber\\
&=&\frac{2}{\pi c_\alpha}\int_{z}^\infty du
\int_{0}^\infty dw \;\cos\left(\frac{w u}{c_\alpha}\right)
\left[E_\alpha(-w^2)\right]^2.\nonumber\\
\end{eqnarray}

The particle density as a function of time is obtained from the interval
distribution function via Eq.~(\ref{c}).
The asymptotic density $\lim_{t\to \infty} c(t)$ then is
\begin{eqnarray}
c(t)&=&-\left. \frac{\partial E^{\mathcal A}(x,t)}{\partial x}\right|_{x=0}  =
2 \mathcal{P}_\alpha(0,t) \nonumber\\
&=& \frac{c_\alpha}{ \sqrt{K_\alpha t^\alpha}} = \frac{z}{x}.
\label{asympt}
\end{eqnarray}
When $\alpha=1$ this reduces to the familiar result $c(t) = (2\pi
Dt)^{-1/2}$.  

At these long times, using the asymptotic portion 
$E^{\mathcal A}(x,t)$ of the solution $E(x,t)$ in Eq.~(\ref{p})
we obtain
\begin{eqnarray}
c(t) p(x,t)&=&  \frac{\partial^2 E^{\mathcal A}(x,t)}{\partial x^2}
=-2  \frac{\partial \mathcal{P}_\alpha(x,t)}{\partial x}\nonumber\\
&=&\frac{2}{\pi} \int_0^\infty dq \; q\; \sin(q x)
\left[E_\alpha(-K_\alpha q^2 t^\alpha)\right]^2 \nonumber\\
&=&\frac{2}{\pi K_\alpha t^\alpha} \int_0^\infty dy \; y\;
\sin\left(\frac{y \,x}{\sqrt{Kt^\alpha}} \right)
\left[E_\alpha(-y^2)\right]^2.\nonumber\\
\label{ctpxt}
\end{eqnarray}
Defining $p(z) dz =p(x,t) dx$ where  $z$ is defined as above, 
we find that for long times
 \begin{equation}
p(z)=\frac{2}{\pi c_\alpha^2} \int_0^\infty dy \; y\;
\sin\left(\frac{z\,y}{c_\alpha} \right)
\left[E_\alpha(-y^2)\right]^2. 
\label{pofz}
\end{equation}

\subsection{Results Valid for All Times for Initial Poisson Distribution}
To find the particle density and interparticle distribution function for
all time requires the solution of the system~(\ref{EcuEa}), which in
turn requires specification of an initial distribution.
We choose an initial Poisson distribution for this analysis,
\begin{equation}
E^{\mathcal T}(x,0)=f(x) = E(x,0) = e^{-c_0x},
\end{equation}
where $c_0\equiv c(0)$ is the initial concentration of particles.
As before, we formulate a related problem,
\begin{equation}
\label{related}
\begin{aligned}
\mathfrak{F}_\alpha \mathcal{E}^{\mathcal T}(x,t)&=0,
\qquad -\infty<x <\infty  \\
\mathcal{E}^{\mathcal T}(-\infty,t)&=0,\\
\mathcal{E}^{\mathcal T}(\infty,t)&=0\\
\mathcal{E}^{\mathcal T}(x,0)&=f(x) =
\begin{cases}
e^{-c_0x}, & x>0 \\
-e^{c_0x}, & x<0.
\end{cases}
\end{aligned}
\end{equation}
By symmetry we deduce that $\mathcal{E}^{\mathcal T}(0,t)=0$, so that
the solution of this problem for $x\ge 0$ is just the solution
$E^{\mathcal T}(x,t)$ of Eq.~\eqref{related}, that is,
$\mathcal{E}^{\mathcal T}(x,t)=E^{\mathcal T}(x,t)$ for $x\ge 0$.
We can write
\begin{equation}
\mathcal{E}^{\mathcal T}(x,0) = \int_{-\infty}^\infty dy
\mathcal{E}(y,0) \delta (x-y).
\end{equation}
Then our knowledge of the Green function for $\mathfrak{F}_\alpha$
and
our hypothesis that this operator is linear immediately leads to the
solution
\begin{equation}
\begin{aligned}
\mathcal{E}^{\mathcal T}(x,t) = &\int_{-\infty}^\infty dy \mathcal{E}(y,0)
\mathcal{P}_\alpha (x-y) \\
=& - \int_{-\infty}^0 dy e^{c_0y} \mathcal{P}_\alpha (x-y)\\
&+ \int_0^\infty dy e^{-c_0y} \mathcal{P}_\alpha (x-y) \\
=& E^{\mathcal T}(x,t) \qquad  \text{for}\quad x \ge 0.
\end{aligned}
\end{equation}
With a bit of manipulation and explicit insertion of Eq.~(\ref{green})
we
finally obtain for $x\ge 0$,
\begin{equation}
\begin{aligned}
E^{\mathcal T}(x,t) = &-2\sinh(c_0x) \int_x^\infty dy e^{-c_0y}
\mathcal{P}_\alpha (y)\\
&+ 2 e^{-c_0x} \int_0^x dy \cosh (c_0y)
\mathcal{P}_\alpha (y).
\label{ea}
\end{aligned}
\end{equation}
The full interval distribution is then the sum of (\ref{Ebxt}) (valid
for
any initial condition and determinative of the asymptotic behavior) and
(\ref{ea}) (explicitly calculated here for a Poisson initial
distribution
and going to zero asymptotically).

As before, the particle density as a function of time is obtained from
the interval distribution function via Eq.~(\ref{c}).  From
Eq.~(\ref{ea}) we calculate
\begin{equation}\label{thesecond}
-\left. \frac{\partial E^{\mathcal T}(x,t)}{\partial x}\right|_{x=0}  =
-2 \mathcal{P}_\alpha(0,t) + 2c_0 \int_0^\infty dy
e^{-c_0y}\mathcal{P}_\alpha(y,t),
\end{equation}
from which upon addition of Eqs.~(\ref{asympt}) and (\ref{thesecond})
and use of Eq.~(\ref{green}) it follows that
\begin{equation}
c(t)=-\frac{2}{\pi \sqrt{K_\alpha t^\alpha}} \int_0^\infty dy
\frac{1}{1+(y^2/c_0^2K_\alpha t^\alpha)} \left[ E_\alpha(-y^2)\right]^2.
\label{ct}
\end{equation}
Note that this result, valid for all times, reduces to (\ref{asympt})
when $t\to\infty$.

The interparticle distribution function follows from Eq.~(\ref{p}). To
add to our previous asymptotic result, we note that
\begin{eqnarray}\label{parEax2}
\frac{\partial^2 E^{\mathcal T}(x,t)}{\partial x^2}&=&2 \frac{\partial
\mathcal{P}_\alpha}{\partial x}
-2c_0^2 \sinh(c_0x)\int_x^\infty dy e^{-c_0 y}
\mathcal{P}_\alpha(y,t)\nonumber\\
&&+2c
_0^2 e^{-c_0 x}\int_0^x dy \cosh(c_0 y) \mathcal{P}_\alpha(y,t).
\end{eqnarray}
Introducing  Eq.~\eqref{green} into \eqref{parEax2}, one finds, after
some manipulations,
\begin{equation}
\begin{aligned}
\label{eafin}
\frac{\partial^2 E^{\mathcal T}(x,t)}{\partial x^2} =& 2 \frac{\partial
\mathcal{P}_\alpha}{\partial x} +
\frac{2}{\pi K_\alpha t^\alpha} \int_0^\infty
\frac{dy}{1+y^2/c_0^2K_\alpha t^\alpha}\\
&\times y \sin\left(\frac{xy}{\sqrt{K_\alpha t^\alpha}}\right)\;
\left[E_\alpha(
-y^2)\right]^2.
\end{aligned}
\end{equation}
Upon addition of this contribution and the asymptotic one obtained
earlier, we finally have
\begin{equation}
\begin{aligned}
\label{ctpxtT}
c(t) p(x,t)=& \frac{2}{\pi K_\alpha t^\alpha} \int_0^\infty
\frac{dy}{1+y^2/c_0^
2K_\alpha t^\alpha}\\
&\times y \sin\left(\frac{xy}{\sqrt{K_\alpha
t^\alpha}}\right)\; \left[E_\alpha(-y^2)\right]^2
\end{aligned}
\end{equation}
This expression reduces to Eq.~\eqref{ctpxt} when $t\to\infty$. Only in
the asymptotic limit is it possible to write $p(x,t)/c(t)$ as a
function of the single combined variable $z=c(t) x$.

\section{A conundrum and some choices}
\label{conundrum}

In the next section we compare our theory to numerical simulation
results and, as we shall see, the comparison is on the whole very
successful.  And
yet we know that the theory is approximate -- indeed, we introduced two
hypotheses that are surely not exact. We note here an additional related
conundrum which exhibits itself (albeit only weakly even when
$\alpha$ is small) in the numerical simulations of the $A+A \to A$ reaction
(but not in the $A+A\to 0$ problem). In the
discrete version of the problem used for the simulations,
a reaction $A+A\to A$
occurs when an $A$ particle steps onto a site already occupied by
another $A$ particle. One then has to decide which of the two particles
is the one that is removed from the system, the one that was there
(``kill" rule) or the one that just stepped onto the site (``no-kill"
rule).  The choice could vary with each reaction event.  In the diffusive problem ($\alpha=1$) the
choice does not matter.  In a subdiffusive situation, however, the choice does matter
since the properties of the subsequent random walk of the survivor, including
the probability that the walker continues to remain at that site, depend
on its age.  In particular, if the survivor is the one that arrived at
the site first then the probability that it will remain at that site
is greater than if the survivor is the new arrival.
The implication is
that the mesoscopic results such as the interparticle distribution
function depend on the microscopic reaction rule.  On the other hand,
we will show that the differences in the observable quantities are small even for $\alpha$
considerably smaller than $1$, but the agreement with simulations is a bit
better using the ``kill"
rule.  This is as expected since this rule resets, so to speak, the ``clock''
of the surviving particle following a reaction event and is thus
in some sense closer to the assumptions inherent in Hypothesis 1.  In most of our simulations we use
the ``kill" rule.

The second choice we must make in our simulations is the form of the distribution of waiting times
between steps.  In most cases we use a Pareto-type waiting time distribution,
\begin{equation}
\psi_P(t) = \frac{\alpha}{t_0(1+t/t_0)^{1+\alpha}},
\label{Paretodistr}
\end{equation}
while in some we use a Mittag-Leffler-based distribution,
\begin{equation}
\psi_{ML}(t) = - \frac{d}{dt} E_\alpha [-(t/t_0)^\alpha]
\label{MLdistr}
\end{equation}
(in our simulations we set $t_0\equiv 1$).
They can both be used to describe
the asymptotic behavior of subdiffusive random walkers. While either form is
mathematically acceptable,
$\psi_{ML}$ is  preferable at short times specially as $\alpha \to
1$~\cite{Masguerra}.
In most of our simulations, times are
long and $\alpha$ is not close to unity, so the choice is not a central issue.
Nevertheless, we address this issue
here even before presenting our simulation methodology and results because
while both of these waiting times do lead to a fractional diffusion equation at long times,
the choice of the waiting time distribution determines the value of the
generalized diffusion coefficient $K_\alpha$.  For the Pareto case $K_{\alpha,P} =
1/2\Gamma(1-\alpha)$, while the Mittag-Leffler form (in our simulation
units) leads to $K_{\alpha,ML} = 1/2$. In the discussion of our results this is the only difference
between theoretical results labeled by $P$ and those labeled by $ML$.  The Pareto generalized
diffusion coefficient diverges as $\alpha\to 1$.  This behavior is symptomatic of other
problems associated with this waiting time distribution in this limit for the calculation of
quantities that explicitly depend on $K_\alpha$, and is a strong motivator for
the introduction of the Mittag-Leffler form.  We will only use the Pareto distribution
in unproblematic regimes of $\alpha$, where these difficulties are not an issue.

\section{Numerical Evidence}  
\label{numerical}

In this section we present numerical simulation results to test the
adequacy of our theory.  Specifically, we present simulation results for
the particle density and for the interparticle
distribution function.  We begin by briefly describing our numerical
simulation methodology.

We proceed via the following steps in our numerical algorithm.  First we
generate a $1d$ lattice.  We then generate an escape time for each particle
chosen from the given waiting time distribution. We choose the particle with the
smallest waiting time.  This particle jumps to one of its two nearest neighbors.
If the destination site is empty, we update the waiting time for the arriving
particle.  If the destination site is occupied, the particles coagulate into a
single particle. This is where we have to specify whether the event is of the
``kill'' variety or of the ``no-kill'' variety, as described earlier.  Next, we
look again for the particle with the smallest waiting time.  The other
particles that have not yet moved simply continue evolving according to
their own internal clocks. The particle that emerges from the
coagulation either starts its waiting time at the moment of the reaction (``kill'') or
continues evolving according to its prior setting, which is unchanged by the
reaction (``no-kill''). The simulation then
continues until it reaches the time $t$ of interest or the concentration $c(t)$
of interest. The experiment is then repeated over and over again for an
ensemble of such chains.

As mentioned earlier, there are issues related to the choice of the waiting
time distribution, specifically whether to use the Pareto form
Eq.~(\ref{Paretodistr}) or the Mittag-Leffler form Eq.~(\ref{MLdistr}).
\cite{Masguerra,Hanggi}. For $\alpha < 1$ they both give the
same asymptotic results. At short times for $\alpha < 1$ and for all times when
$\alpha=1$ the
Mittag-Leffler distribution is the appropriate one to use. Most of the results
presented below are asymptotic and for $\alpha<1$, and we mostly use the Pareto
distribution, having
ascertained to our satisfaction that both lead to the same outcome.  For
short-time results we use the Mittag-Leffler distribution. 
Note that when looking at short-time behavior we need to specify the initial
distribution of particles over the line.  We have consistently chosen a random
Poisson initial distribution.

\subsection{Results}
\begin{figure}
\begin{center}
\includegraphics[width=1.0\columnwidth]{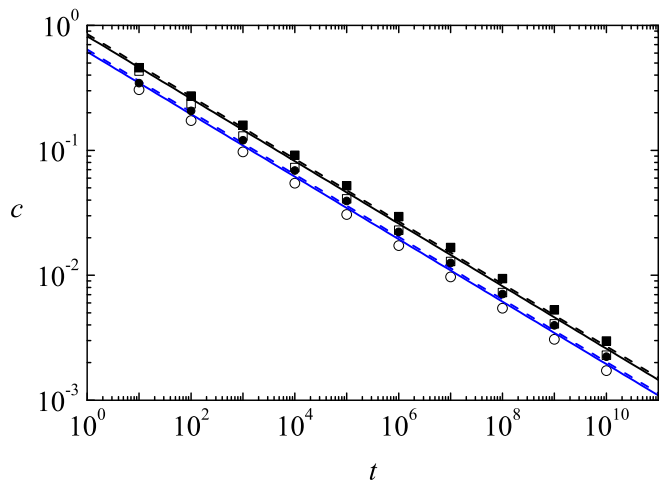}
\end{center}
\caption{Reactant concentration vs time on a lattice of $10^4$ sites with periodic boundary
conditions.  $c(0)=1$, $\alpha=0.5$.  Symbols give results of numerical simulations with kill (open
symbols) and no-kill (black symbols) protocols.  Circles: Mittag-Leffler-based
waiting time distribution. Squares: Pareto waiting time distribution.  Solid lines: our theory,
Eq.~(\ref{ct}), with $K_{alpha,P}$ (upper line) and with $K_{\alpha,ML}$ (lower line). Broken lines:
approximations $c(t)\sim 1/S_P(t)$ (upper)  and $c(t)\sim 1/S_{ML}(t)$ (lower).  For a definition of
symbols and for quantitative fits to the various lines see text. In addition,
in all the figures of this paper the error bars are smaller than the
symbol sizes.}
\label{fig1}
\end{figure}

We will now test our theory against simulation results, and anticipate that the agreement
is gratifyingly good. Figure~\ref{fig1} contains a variety of results for the concentration $c(t)$
of surviving $A$ particles as a function of time for $\alpha=0.5$ and an initial
concentration $c(0)=1$.  The
initial concentration is sufficiently high for there to be essentially no transient behavior before
the asymptotic power law dependence of $c(t)$ takes over, as evidenced by the
straight lines.
First, note the comparison between the simulation (\textit{sim}) results with a
Pareto-like waiting time
distribution (squares) and those of a Mittag-Leffler form (circles). Both are essentially linear, as
expected.  The slopes are the same, but the Pareto results are a little higher.  A best fit leads to
$c_{P,sim}\sim 0.751~ t^{-0.252}$ and $c_{M,sim}\sim
0.541~ t^{-0.250}$.
The open symbols are ``kill" simulations (and the fits just given are for this
case), while the
solid symbols are for ``no-kill."  The ``no-kill" rule leads to consistently higher concentrations.
This makes sense since the survivor at each reaction event is the $A$ that arrived first at the
reaction site, and it remains there longer (due to aging) than would the other reaction partner.
This leads to its longer survival.

While these results and comparisons are interesting, our most important task is to compare these
results with those of our theory, which is shown by the upper solid line for the Pareto case and the
lower solid line for the Mittag-Leffler case.  The slopes are in excellent agreement with those of
the simulations.  The coefficients fall precisely between the ``kill'' and
``no-kill'' results.
Specifically, we find that $c_{P,theory} \sim 0.819  ~t^{-0.25}$ and $c_{ML,theory}\sim
0.615~ t^{-0.25}$.  
One might be tempted to conjecture that the theoretical model in some
sense lies between the ``kill" and ``no-kill" scenarios.  For example,
perhaps the model is most germane if ``kill'' or ``no-kill'' are randomly
selected according to some appropriate distribution. For now, this remains
a matter of conjecture.

Finally, one additional result is shown in Fig.~\ref{fig1}, namely, the validity
of the relation
between the concentration of surviving reactant and the distinct number of sites visited by a random
walker, $c(t)\sim 1/S(t)$~\cite{MetKlaPhysRep,Yustebook}.  The dashed lines show
$1/S(t)$, with the upper dashed line corresponding
to the Pareto case, $1/S_P(t)\sim 0.853~  t^{-0.25}$, and the lower dashed line associated with
the Mittag-Leffler choice, $1/S_{ML}(t)\sim 0.641~ t^{-0.25}$.

\begin{figure}
\begin{center}
\includegraphics[width=1.0\columnwidth]{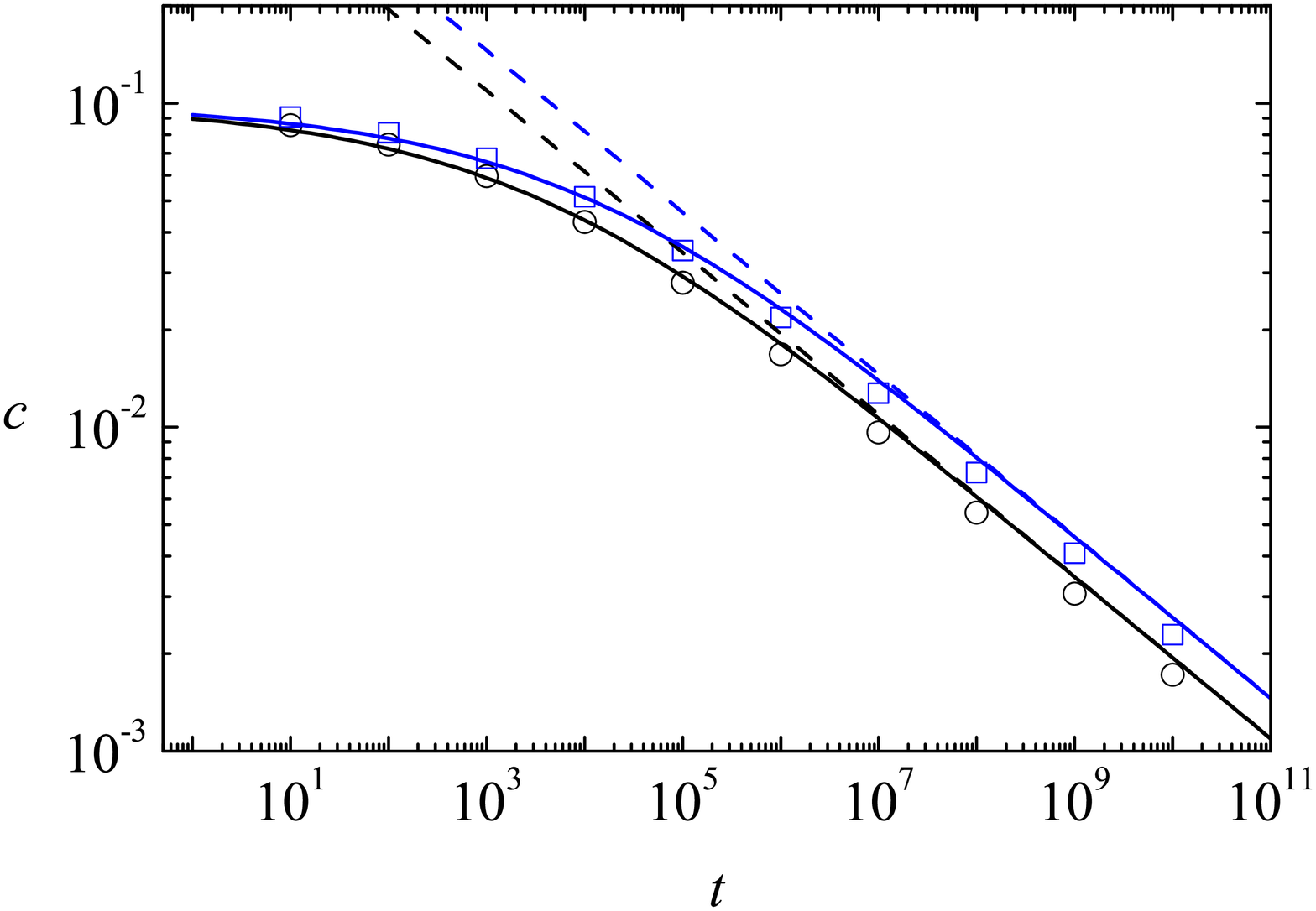}
\end{center}
\caption{Reactant concentration vs time on a lattice of $10^4$ sites with periodic boundary
conditions.  $c(0)=0.1$, $\alpha=0.5$, showing transient behavior.
Symbols give results of
Numerical simulations with ``kill" protocol for Pareto (squares) and
Mittag-Leffler-based (circles) waiting time distribution. Solid lines: our theory,
Eq.~(\ref{ct}), with $K_{\alpha,P}$ (upper line) and $K_{\alpha,ML}$ (lower
line). Broken lines:
asymptotic approximation for $c(t)$ as given in Eq.~(\ref{asympt}) for both cases.}
\label{fig2}
\end{figure}
Figure~\ref{fig2} again shows the concentration of surviving reactant as a function of time for
$\alpha=0.5$ on a lattice of $10^4$ sites, but now we explore whether our full
theory, Eq.~(\ref{ct}),
in fact also captures the transient behavior.  We explore this behavior by
starting with a lower
initial concentration, $c(0)=0.1$.  Clearly, the theory again works extremely well for all times.
Here the simulations are carried out in the ``kill" scenario only, and we have presented both the
Pareto (squares) and Mittag-Leffler (circles) results.  The solid curves are the theoretical results
for the Pareto (upper) and Mittag-Leffler (lower) cases. The dashed lines are the asymptotic results
Eq.~(\ref{asympt}) for the two cases.

We stress that the asymptotic exponent of time as predicted by all theories (including our earlier
theory that suffers from other difficulties) is correct and agrees with simulation results,
which also agree with each other. The different theories and simulations (Pareto vs Mittag-Leffler,
``kill" vs ``no-kill," our earlier theory vs our current theory, distinct number of sites visited predictions)
lead to different (but not wildly different) prefactors.  A more stringent test is provided by the
interparticle distribution function, for which we now present a series of figures.

\begin{figure}
\begin{center}
\includegraphics[width=0.7\columnwidth]{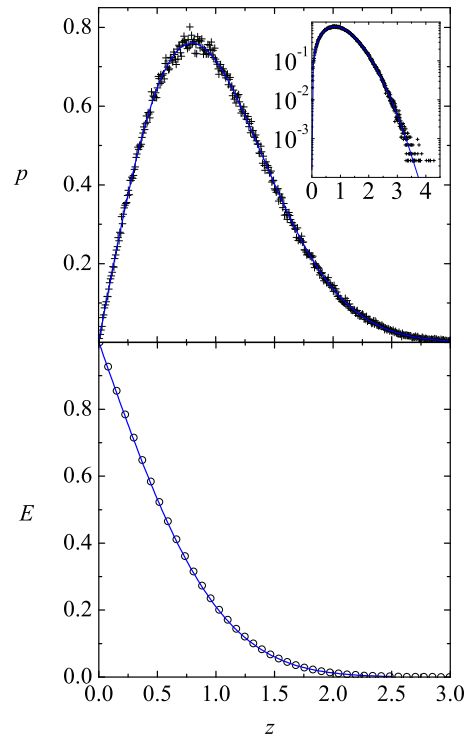}
\end{center}
\caption{Asymptotic interparticle distribution function and empty interval function vs the scaled variable
$z=c(t)x$ for the classic reaction-diffusion problem.  The simulation results (symbols) are
obtained with an exponential waiting time distribution on a lattice of 40,000 sites at time
$t=6000$. The solid curves from the traditional exact theories are given in Eqs.~(\ref{classicp})
and (\ref{classicE}). The upper panel shows the interparticle distribution function and the inset shows
it on a logarithmic scale that exhibits the simulation scatter at extremely low densities.  The lower
panel shows the empty interval function.}
\label{fig3}
\end{figure}

In the (``normal") reaction-diffusion problem the interparticle distribution function develops a growing gap
at small distances.  This gap arises as spatially close pairs react and are not replenished because
diffusion is slow in
one dimension. The gap explains the ``anomalous" decay law $c(t) \sim t^{-1/2}$, called
anomalous because it differs from the law of mass action behavior $c(t)\sim t^{-1}$ predicted
for a Poissonian distribution of well-mixed reactants.  Figure~\ref{fig3} shows these results
as obtained from theory and simulations. They agree extremely well, which is a confirmation that we
are in the asymptotic regime at time $t=6000$. The exact theoretical expressions
are well known~\cite{Torney,Spouge},
\begin{equation}
 p(z) = \frac{\pi}{2} z e^{-\pi z^2/4}
\label{classicp}
\end{equation}
and
\begin{equation}
 E(z)={\rm erfc} (\sqrt{\pi} z/2).
\label{classicE}
\end{equation}
This figure serves as a basis of comparison for subsequent subdiffusive results.

\begin{figure}
\begin{center}
\includegraphics[width=0.7\columnwidth]{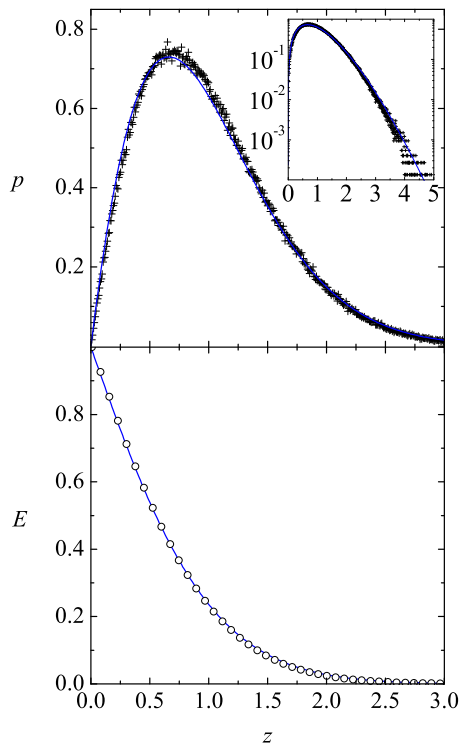}
\end{center}
\caption{Asymptotic interparticle distribution function and empty interval function as a function
of the scaled variable $z=c(t)x$ for $\alpha=0.7$.  The simulation results (symbols) are obtained
with a Pareto waiting time distribution and a ``kill" rule on a lattice of 40,000 sites at
time $t=10^6$. The solid theoretical curves are obtained from Eqs.~(\ref{pofz}) and (\ref{Ez}).
The upper panel shows the interparticle distribution function and the inset shows it on
a logarithmic scale that exhibits the simulation scatter at extremely low densities.  The
lower panel shows the empty interval function.}
\label{fig4}
\end{figure}

Figure~\ref{fig4} shows the same three panels for the subdiffusive case $\alpha=0.7$, but it
is necessary to go to longer times, $t=10^6$, to arrive at the asymptotic behavior.  The theoretical
curves are obtained from Eqs.~(\ref{pofz}) and (\ref{Ez}) and the simulations are carried out using
a Pareto waiting time distribution with a ``kill'' rule. The agreement between
theory and simulations is
still excellent, with very small differences apparent near the maximum of the interparticle distance
distribution. We have ascertained that we are in the asymptotic regime, so these differences
are an indication of the approximate nature of the theory rather than of uncertainties
pre-asymptotic transient effects.
\begin{figure}
\begin{center}
\includegraphics[width=0.7\columnwidth]{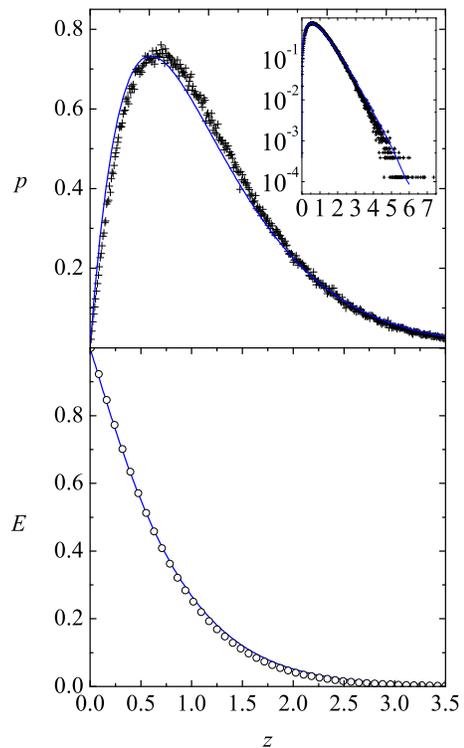}
\end{center}
\caption{Asymptotic interparticle distribution function and empty interval function as a function
of the scaled variable $z=c(t)x$ for $\alpha=0.2$.  The simulation results (symbols) are obtained with a
Pareto waiting time distribution and a ``kill" rule on a lattice of 40,000 sites at time $t=10^{19}$. The
solid theoretical curves are obtained from Eqs.~(\ref{pofz}) and (\ref{Ez}). The upper panel
shows the interparticle distribution function and the inset shows it on a logarithmic scale that
exhibits the simulation scatter at extremely low densities.  The lower panel shows the empty interval function.}
\label{fig5}
\end{figure}
Figure~\ref{fig5} shows the three panels once again, now for the extremely subdiffusive case
$\alpha=0.2$.  The particles spend a great deal of time simply waiting between jumps, and so it
is necessary to go to much longer times, here $t=10^{19}$, to reach asymptotic behavior. The differences
between theory and simulation are still small but certainly more noticeable. It is of course not clear whether
the differences arise from the fact that the distribution of particles is affected by the reaction, or from
the assumption of linearity of the evolution operator of the empty interval function, or both.
In any case,
it does not seem an exaggeration to assert that the theory is very good.  It is
also clear
that a gap in the interparticle distance distribution develops no matter how subdiffusive the
system, contrary to our earlier thinking~\cite{ourold}.

\begin{figure}
\begin{center}
\includegraphics[width=0.7\columnwidth]{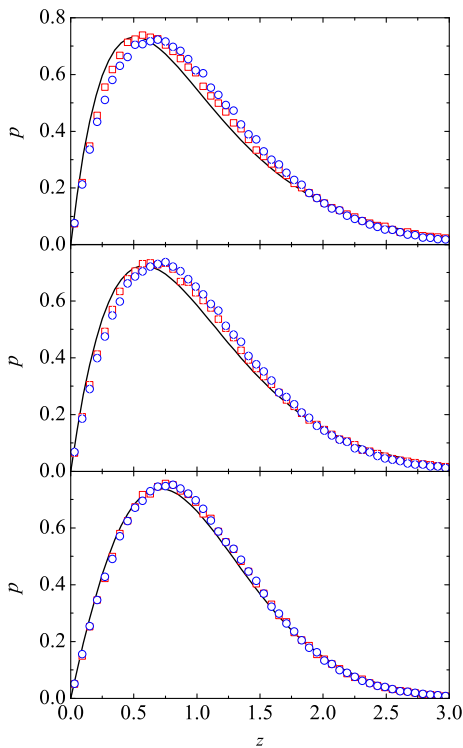}
\end{center}
\vspace*{0.25in}
\caption{Interparticle distribution function $p(z)$.
Symbols: simulation results with a Pareto waiting time distribution.  Upside
down triangles (red in color): ``kill"
rule. Right side up triangles (blue in color): ``no-kill" rule.  Lattice size
$L=10000, c(0)=1, c(t_{final})=0.0025$.
Upper panel: $\alpha=0.2$. Middle panel: $\alpha=0.5$. Lower panel:
$\alpha=0.8$. The solid curves are calculated from Eq.~(\ref{pofz}).}
\label{fig6}
\end{figure}

Two further issues are addressed in the following figures. In most of our simulation results
we have used the ``kill" rule whereby the walker that arrives at an already occupied site eliminates
the particle that was there, and we have stated that agreement with our theory is better
with this rule than with the ``no-kill" rule whereby the newcomer is eliminated.  Figure~\ref{fig6}
shows three panels where this is illustrated for $\alpha=0.2$, $0.5$, and $0.8$ (the rule
choice does not matter when $\alpha=1$).  We see that as $\alpha$ decreases the differences in
the simulation results for these two cases increase, and that the theory is closer to the
``kill" results. Note that in these figures instead of a final simulation time we report a
final simulation concentration to take advantage of the fact that $z$ depends on time only through
the concentration and hence we avoid additional uncertanities.
These differences point to the approximation inherent in the hypothesis that the
reaction
does not alter the spatial distribution of reactants.  It does, although not by very much, especially
if $\alpha$ is not extremely small.

\begin{figure}
\begin{center}
\includegraphics[width=0.7\columnwidth]{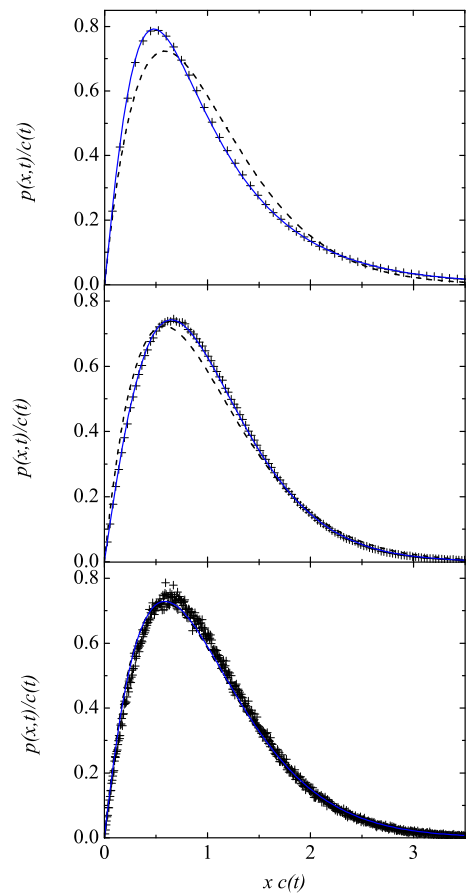}
\end{center}
\caption{ Interparticle distribution function for $\alpha=0.5$.
Symbols: simulation results with a Mittag-Leffler waiting time distribution with
$C(0)=0.1$.
Upper panel: $t=10^2$. Middle panel: $t=10^5$. Lower panel:
$t=10^8$. The solid curves are asymptotic, calculated from Eq.~(\ref{pofz}).}
\label{fig7}
\end{figure}

Finally, we briefly examine the approach of the interparticle distribution to its asymptotic behavior by
plotting $p(x,t)/c(t)$ as a function of $xc(t)$, the scaling variable.  Figure~\ref{fig7} shows the
time progression of the distribution for $\alpha=0.5$ until its arrival at asymptotic behavior in
the lower panel.  It is preferable to use the Mittag-Leffler waiting time distribution for this progression as we
have done in these simulations because of its advantages at early times compared to the Pareto distribution.
The figure illustrates that the growth of the interparticle gap at early times is faster than
at later times when it is determined by the scaling form, but that the system settles into its
asymptotic form rather quickly.

\section{Conclusions}
\label{conclusions}

We have presented a new theory for the coagulation reaction $A+A \to A$ on a
lattice where the $A$'s perform a subdiffusive continuous time random walk
characterized by the subdiffusive exponent $\alpha$.
Our theory relies on the connection between continuous
time random walks and fractional diffusion equations, and is based on two
assumptions. The first hypothesis is that the evolution of the probability
density for an empty interval of
length $x$ at time $t$ in the presence of
reactions is the same as the probability density that, in the absence of
reactions, the distance between two
subdiffusive particles that start at the same location at $t=0$ is $x$ at time
$t$. The two probability densities are not equal because they obey
different initial and boundary conditions.
Only the evolution equations are assumed to be the same. The
second hypothesis is that the (unknown) common equation of evolution for these
two probability densities is linear.  We have no analytic way of checking the
validity of these assumptions, but the results of the theory are extremely close
to those of numerical simulations in all cases tested.  Some of the small
differences are more likely ascribed to the second hypothesis, the
assumption of linearity of the evolution operator, than the first, as explained
in the earlier analysis.

We explicitly calculated the global reactant concentrations as a function of
time.  This is not a very discriminating measure of the quality of different
theories; all the theories and implementations tested in this work lead to the
same correct decay exponent of time, $c(t)\sim t^{-\alpha/2}$, and only the
prefactors vary, albeit not dramatically.  The well known
association of $c(t)$ with the inverse of the distinct number of sites visited
by a subdiffusive walker also gives the correct exponent and a prefactor within
the
range of our theory and simulation results. We also tested the early time
transient behavior of $c(t)$ predicted by our theory and find that 
agreement with simulation results is also excellent. 

A more stringent test of the theory is the interparticle distribution function,
that is, the more ``local'' function $p(x,t)$.  Again, we find our comparisons
to show that the theory captures the simulation results very well, even
for extremely subdiffusive systems.  Most of the results that we present are
asymptotic, but we also tested the validity of our theory in the transient
regime, and found again that it works equally well. 

An issue that we discussed at some length concerns a choice that must be made at
the microscopic (continuous time random walk) level used in the simulations but
not at level of the mesoscopic fractional diffusion equation used in the theory.
The
choice concerns the particle that vanishes when two $A$ particles land at the
same location, the one that was there already or the one that just arrived. The
results of simulations show the differences to be small but discernible in one
dimension,
which is evidence of aging~\cite{Barkai}; we find that the agreement with
our theory is
slightly better with the first choice and perhaps would be even better with a
combination of choices that allows for both possibilities with some probability
distribution.  We also discussed the choice of the waiting time distribution, an
issue that becomes particularly important at early times and when $\alpha$
approaches unity.

Finally, we point to our work in higher dimensions~\cite{ourlatest}.  In that
work we considered the same reaction as well as the annihilation reaction but
on a lattice with traps of random depths.  The depths of the traps are
associated with mean sojourn times that are finite and distributed according to
a power law. Using a mean field formulation, we approximated this system by one
that has identical fat-tailed waiting time distributions for each site, and
show that in dimension $d=3$ the mean field model already captures
the reaction dynamics and distribution functions of the original random trap
system.  In the mean field model, the ``kill" and ``no-kill" scenarios give
the same results.
A particularly important result in that work is that the waiting time
distribution is unaffected by the reaction. In our one-dimensional model we
start with the translationally invariant system, but also postulate, and show
to lead to excellent results, an equivalence between evolution operators
in the systems with and without reactions.
We also point to the fact that the distribution develops a growing
gap between particles~\cite{ourold}.  The growth of the interparticle gap at
early times is faster than at later times when it is determined by a scaling
form, but in any case there is a gap.  This ``anomalous'' distribution leads to
the ``anomalous'' decay kinetics of the concentrarion, as in the case
of normal diffusion in one dimension.

The analysis presented here can be adapted to the $A+A\to 0$ reaction, but
the quantities to be calculated there are different because gap lengths don't
change continuously.  One concentrates instead on the parity of the number of
particles in a given interval~\cite{Something2,Something3}, from which the
desired quantities can in turn be derived.  This will be a subject for future
work.

\acknowledgments     
This work was partially supported by the Ministerio de Ciencia y
Tecnolog\'{\i}a (Spain) through Grant No. FIS2007-60977, by the Junta de
Extremadura (Spain) 
through Grant No. GRU09038, and by the
National Science Foundation under grant No. PHY-0354937.

\end{document}